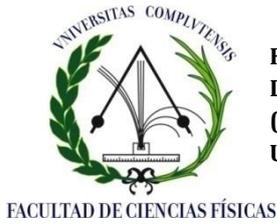

**Facultad de CC Físicas.**
**Dpto. Física de la Tierra, Astronomía y Astrofísica I**
**(Geofísica y Meteorología)**
**Universidad Complutense de Madrid**

*Trabajo Fin de Grado:*

**ESTRUCTURA Y DINÁMICA PLANETARIA EN LA CIENCIA FICCIÓN.**

**ANÁLISIS Y CRÍTICA CIENTÍFICA DE *INTERSTELLAR.***

**Autor: Rodrigo González Peinado**

**Fecha: 22/06/2015**

*"Do not go gentle into that good night,*
*Old age should burn and rave at close of day;*
*Rage, rage against the dying of the light.*
*Though wise men at their end know dark is right,*
*Because their words had forked no lightning they*
*Do not go gentle into that good night.*
*Rage, rage against the dying of the light"*

Dylan Thomas (1914-1953)

# Agradecimientos

Me gustaría agradecer a la profesora María Luisa Osete López del Departamento de Física de la Tierra, Astronomía y Astrofísica I (Geofísica y Meteorología) y al profesor Javier Gorgas García del Departamento de Física de la Tierra, Astronomía y Astrofísica II (Astrofísica y Ciencias de la Atmósfera) de la Universidad Complutense de Madrid, su inestimable colaboración en la realización de este trabajo. Soy consciente de que sin su ayuda y conocimientos, no habría sido capaz de realizar este trabajo. A su vez, quería mostrar mi gratitud por su tiempo y dedicación, los cuales me han servido de punto fuerte de apoyo.

Mención también a la Facultad de Ciencias Físicas de la Universidad Complutense de Madrid, ya que me proporcionó los medios y las referencias necesarias para el desarrollo de este trabajo, en especial, al personal no docente del Departamento de Física de la Tierra, Astronomía y Astrofísica I (Geofísica y Meteorología).

Además, me gustaría agradecer también a las numerosas instituciones a nivel global, autores, revistas, etc. la gran cantidad de información que prestan para uso público y con el único objetivo de hacer que la Ciencia en general, y la Física en particular, sean de bien común

Por último, agradecer a mis familiares y amigos su insistencia con *Interstellar*, ya que me dieron muchas ideas que plantear en este trabajo. Espero que sus dudas queden resueltas al finalizar esta memoria.

I

# Summary

> *"Science is about admitting what we don't know"*
> Murph

Christopher Nolan's latest blockbuster, *Interstellar*, has supposed a revolution not only from a cinematographically viewpoint, but also in the relation between current spectators and science. The aim of this report is to analyze some features presented in *Interstellar*. Basic Newtonian Physics shows how a planet orbiting a supermassive black hole like Gargantua must travel at 0.25 the speed of light in order to be in a stable orbit. Thus, this orbit is too far from the black hole to explain the time dilatation observed in Miller's Planet and an the orbit that could explain such a time dilatation is much more smaller that Roche's limit. Furthermore, I explain how enormous waves as Miller's Planet ones could appear by studying the tidal potential created by Gargantua, obtaining a wave of 200 kilometers high, but better results are obtained by considering a series of tsunamis in Miller's Planet. Finally, I present an explanation to Mann's Planet ice-covered landscape in which *Snowball Earth Theory* plays a fundamental role. No difficult General Relativity formulas will be used in order to make easier the comprehension of the research.

*Spoiler Alert: this study contains references to the film's plot.

# Índice

# Objetivos e introducción

> *"Nature is not evil. Formidable. Frightening. But…no, not evil"*
> Amelia Brand

Los principales objetivos de este trabajo son:

- Analizar científicamente alguno de los supuestos que aparecen en *Interstellar*, así como su comprensión.
- Estudiar los posibles errores que se observan en la película.
- Proponer mecanismos que solventen los problemas o refuercen las hipótesis de partida, aplicando los conocimientos adquiridos durante el Grado en Ciencias Físicas.

Con el deseo de cumplir los objetivos anteriormente mencionados, la estructura del trabajo será la siguiente:

Primero, realizaré una introducción a agujeros negros, explicando su naturaleza y su comportamiento, seguido de una introducción a la dilatación gravacional temporal y a la teoría de las mareas.

A continuación, se presentarán las evidencias empíricas sobre un agujero negro supermasivo en el centro de la Vía Láctea, así como diferentes manifestaciones de las mareas en el Sistema Solar, partiendo desde la Tierra y considerando los casos de los satélites Ío, Europa y Encélado.

El siguiente paso será aplicar estas nociones a *Interstellar*. Para ello, nos centraremos en la naturaleza del agujero negro supermasivo, Gargantúa, y nos desplazaremos hacia sus afueras, estudiando de qué manera éste afecta a la órbita del planeta de Miller y a las olas que en él se producen. Por último, daremos una posible explicación a encontrar un planeta totalmente cubierto de hielo como es el planeta de Mann.

Argumento de *Interstellar*: la Tierra se ha quedado sin cultivos. La única forma que tiene la humanidad para sobrevivir pasa por abandonar la Tierra y buscar un planeta habitable en donde empezar de nuevo. En esta aventura, el capitán de la misión, Cooper (interpretado por Matthew McConaughey), junto con su tripulación, Amelia Brand (Anne Hathaway), Romilly (David Gyasi), Doyle (Wes Bentley), su inseparable robot TARS, y motivados por el profesor Brand (Michael Caine) se enfrentarán a los peligros del viaje interestelar y viajarán a planetas inexplorados con el fin de dar a la humanidad una última oportunidad para la supervivencia.

# 1. Fundamentos teóricos y metodología

*"I'm not afraid of death. I'm an old physicist. I'm afraid of time"*
Dr. Brand

Gran parte del éxito cosechado por *Interstellar* se debe al deseo de Christopher Nolan de ser lo más fiel posible a la Física de verdad. Para ello, contó con la colaboración de uno de los mayores expertos en Cosmología del mundo, el profesor Kip Thorne. En este capítulo se presentan algunos conceptos sobre agujeros negros, dilatación del tiempo y mareas que serán de utilidad para entender *Interstellar.*

## 1.1. Agujeros negros, agujeros negros supermasivos y *wormholes*

Un agujero negro es un objeto astrofísico compuesto únicamente de espacio y tiempo curvos, es decir, los agujeros negros no se componen de materia propiamente dicha. En 1916, Karl Schwarzschild encontró una solución a las ecuaciones de Einstein para un cuerpo masivo no rotatorio y simétricamente esférico. Esta solución establece que existe una superficie, denominada horizonte de eventos, por debajo de la cual nada que lo sobrepase pueda escapar. El horizonte de eventos viene determinado por el radio de Schwarzschild, que toma la forma (Wald, 1984):

$$R_S = \frac{2GM}{c^2} \qquad (1)$$

en donde G es la Constante de Gravitación Universal ($G = 6.67 \cdot 10^{-11}\ N \cdot m^2 \cdot kg^{-2}$), M es la masa del cuerpo y c es la velocidad de la luz ($c = 3 \cdot 10^8\ m \cdot s^{-1}$). A cada cuerpo se le asocia un radio de Schwarzschild que depende únicamente de su masa. Por definición, a todo objeto cuyo radio sea menor a su radio de Schwarzchild se le conoce como agujero negro.

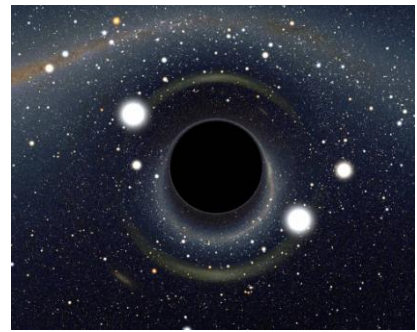

Fig.1.Simulación realizada por ordenador en la que se aprecia cómo la luz de las estrellas se curva en presencia del agujero negro. Imagen de Alain Riazuelo (2010) vía http://apod.nasa.gov/apod/ap101207.html

Como un agujero negro no deja escapar nada de lo que atraviese su horizonte de eventos, ni siquiera la luz, es imposible que sea observado de forma usual. Por ello, la única forma de detectarlo es mediante el efecto óptico que éste produce sobre su alrededor, curvando la luz de las estrellas o galaxias que están detrás de él. Este efecto se conoce como lente gravitacional (Figura 1) y juega un papel muy importante en el desarrollo del agujero negro de *Interstellar*, Gargantúa.

Los agujeros negros se forman a partir del colapso de estrellas supermasivas (con masas decenas de veces la masa del Sol) en los últimos momentos de sus vidas, cuando no les queda combustible por fusionar en su núcleo.

Un caso extremo de agujeros negros son los agujeros negros supermasivos o SMBH por sus siglas en inglés (*Supermassive Black Holes*). Estos agujeros negros poseen una masa asociada de miles a cientos de miles masas solares y constituyen uno de los principales objetos de estudio en la Astrofísica actual. Se cree que se forman por fusión de muchos agujeros negros menores, aunque esta hipótesis es discutida. Más aceptada es la idea de que existe uno de estos SMBH en el centro de cada galaxia, alrededor del cual giran todos los componentes de una galaxia.

Existe un tercer tipo de agujeros negros denominados *wormholes* o agujeros de gusano, propuestos por Flamm en 1916. Según las ecuaciones que gobiernan la Relatividad General, podría existir un pasadizo interdimensional que comunique dos puntos separados del espacio-tiempo (Figura 2). A ese túnel se le dio el nombre de agujero de gusano y se situaría en una dimensión superior a la nuestra llamada *bulk*. Aunque es verdad que las ecuaciones de Einstein predicen su existencia, todavía no se han hallado pruebas concluyentes de que se den en el universo.

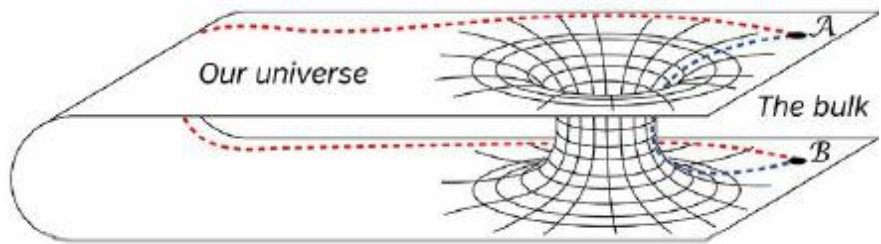

Fig. 2. Agujero de gusano de Flamm. Se aprecia como el agujero de gusano conecta los puntos A y B de nuestro universo (en 2D) a través de un túnel situado en el *bulk*. Imagen de Thorne (2014).

El estudio de los agujeros negros y su naturaleza ha sido un tema de gran interés para la Cosmología Moderna ya que en ellos podría estar la clave para resolver el gran problema de la Física actual: unificar la Relatividad General de Einstein con el Modelo Estándar de Física de Partículas.

## 1.2. Dilatación gravitacional del tiempo

La dilatación gravitacional del tiempo es una consecuencia directa de la Teoría de la Relatividad General de Einstein. Según esta teoría, el tiempo transcurre más lentamente cerca de objetos muy masivos, es decir, el tiempo se dilata a medida que nos acercamos a cuerpos de alta masa. La dilatación gravitacional del tiempo no es única, sino que depende de la distribución de masa del cuerpo y de cómo ésta influye en la geometría del espacio-tiempo de su alrededor. Así, la dilatación gravitacional del tiempo que sufre un observador en las proximidades de un agujero negro simétricamente esférico y no rotatorio (agujero negro de Schwarzschild) viene dada por:

$$\Delta t = \frac{\Delta t_0}{\sqrt{1 - \frac{R_s}{R}}} \quad (2)$$

en donde $\Delta t_0$ es el intervalo de tiempo propio del observador, $R_s$ es el radio de Schwarzschild proporcionado por la expresión (1) y $R$ es la distancia del observador al agujero negro. Nótese cómo a medida que un observador se acerca al radio de Schwarzschild, el intervalo de tiempo entre dos sucesos aumenta, por lo que el tiempo transcurre más despacio, es decir, se dilata.

### 1.3. Mareas

Las mareas son eventos en los que la forma de un cuerpo, más o menos elástico, se ve alterada por la presencia de otro. Estos procesos ocurren a todas las escalas, pero sus consecuencias son mayores cuanto mayores sean las masas de los cuerpos involucrados y cuánto menor sean las distancias entre ellos. Debido a esto, el fenómeno de las mareas se estudia principalmente a nivel planetario. Para entender mejor cómo funciona el fenómeno de las mareas, nos situaremos en el contexto más familiar para nosotros, el sistema Tierra–Luna.

Las mareas en la Tierra se producen por el efecto combinado del Sol y de la Luna sobre la superficie de la Tierra. Este efecto hace que, tanto la masa oceánica como la terrestre, se desplacen de sus posiciones de equilibrio.

El causante de este desplazamiento es el potencial de marea, que, en primera aproximación, toma la siguiente forma (Torge, 1991):

$$U_L = \frac{2}{3} G(r) \cdot \left(\frac{c}{R}\right)^3 (3\cos^2\theta - 1) \quad (3)$$

en donde c es la distancia media entre la Tierra y la Luna, R es la distancia entre los centros de los astros, θ es la latitud del punto del observador en la Tierra y $G(r)$ es la Constante Geodésica de Doodson, cuyo valor es (Torge, 1991):

$$G(r) = \frac{3}{4} G M_L \left(\frac{r^2}{c^3}\right) \quad (4)$$

en donde G es la Constante de Gravitación Universal y r es la distancia desde el centro de la Tierra al punto del observador. $M_L$ es la masa del cuerpo que crea el potencial, en este caso, de la Luna. El potencial de marea que crea el Sol sobre la Tierra es análogo a la expresión (3) cambiando los parámetros lunares por los solares. El desarrollo físico para obtener el potencial de marea puede encontrarse en Torge (1991).

La distancia que separa la superficie en equilibrio de la deformada por el potencial de marea recibe el nombre de altura de la marea, y depende del valor de la gravedad del planeta. En módulo, su expresión es (Torge, 1991):

$$\zeta = \frac{U_L}{g} \qquad (5)$$

en donde *g* es el módulo de la aceleración de la gravedad en la Tierra (*g=9.8 m·s$^{-2}$*).

El potencial de marea puede descomponerse en tres sumandos dependiendo de cómo varíe cada término con el ángulo horario (H) del cuerpo que produce el potencial (Melchior, 1966). Esta descomposición recibe el nombre de descomposición de Laplace. Los tres términos (o tipos de mareas) son:

- $U^0$: periodicidad mixta. Este término sólo depende de la latitud y de la declinación del punto que sufre la fuerza de marea en la Tierra (no tiene dependencia con el ángulo horario). Corresponde a una función en armónicos esféricos zonales, en la que las líneas nodales son paralelos. Posee una periodicidad semimensual en el caso de la Luna (14 días) y semianual en el caso del Sol (6 meses). El potencial es máximo en el ecuador, lo que produce mareas altas, mientras que es mínimo en los polos, donde se producen mareas bajas. Esta variación en la distribución de masas resultante se traduce en un aumento del momento de inercia de la Tierra.
- $U^1$: periodicidad diurna. La dependencia con el ángulo horario es de la forma cos(H) y corresponde a una función teseral. Presenta, como líneas nodales, dos meridianos y un paralelo, coincidente con el ecuador. Debido a la presencia de estas líneas nodales, las regiones en las que se divide la esfera cambian de signo con la declinación del cuerpo que genera el potencial. La amplitud de la marea alcanzará su máximo cuando la latitud sea de 45º (Norte o Sur), que es cuando la declinación del cuerpo es máxima. Las variaciones en la distribución de masas en la superficie de la Tierra causan una variación en la posición del eje principal de inercia.
- $U^2$: periodicidad semidiurna. Presenta una dependencia con el ángulo horario de la forma cos(2H) y sus líneas nodales son meridianos que dividen a la esfera en cuatro regiones, por lo que se trata de una función sectorial. El potencial en la esfera es alternante, habiendo dos sectores con potencial positivo y dos con potencial negativo. Las zonas de potencial positivo corresponderán a mareas altas, mientras que las regiones de potencial negativo corresponden a mareas bajas. Es el término dominante del potencial de marea. Este efecto no produce variación ni en la posición del momento de inercia ni en su valor.

Las mareas producidas por el Sol representan el 46% de las mareas en la Tierra. Por tanto, la principal causa de las mareas en la Tierra es la Luna (Torge, 1991). Una representación de las tres componentes principales de la marea puede verse en la Figura 3.

TIPOS DE MAREA

Periodicidad mixta      Periodicidad diurna      Periodicidad semidiurna

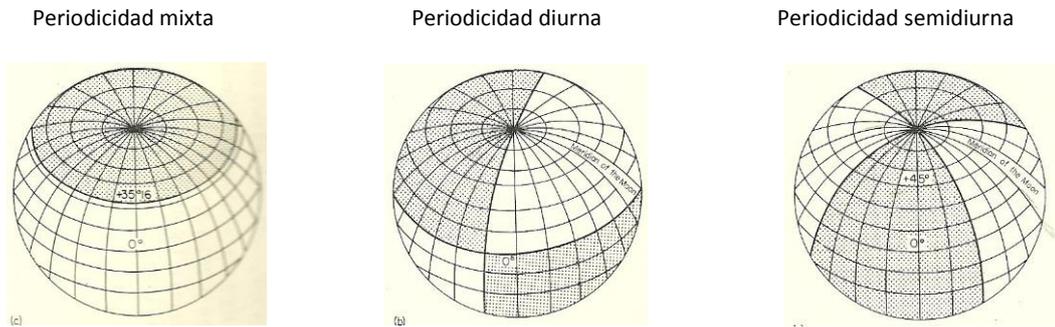

Fig. 3. Representación de las tres principales mareas según su periodicidad en armónicos esféricos. Imagen de Melchior (1966)

Cabe destacar que el efecto de las mareas no sólo afecta a la masa de agua del planeta. Existe otro tipo de marea, denominada marea terrestre, en la que la litosfera se deforma de igual manera que lo hacen los océanos, aunque con unas variaciones mucho menores que las del agua. Este proceso viene determinado por unos coeficientes denominados coeficientes de Love y Shida.

Como ya se ha visto, las mareas son unas fuerzas que tienden a deformar a los cuerpos que las padecen. El caso extremo se presenta cuando un cuerpo que siente las fuerzas de marea se desintegra como consecuencia de las mismas. A la distancia a la que un cuerpo ya no puede mantenerse gravitacionalmente y se rompe debido a las fuerzas de marea se le denomina límite de Roche (Figura 4). El límite de Roche depende de la composición y propiedades de cada cuerpo. Si estudiamos el límite de Roche en sistema Tierra-Luna y consideramos que la Luna se desintegra debido a las fuerzas de marea que genera la Tierra, podemos encontrar dos casos contrapuestos. Para un Luna totalmente rígida, el límite de Roche viene dado por (Lowrie, 2007):

$$d_R = 1.26 R_T \left(\frac{\rho_T}{\rho_L}\right)^{1/3} \qquad (6)$$

en donde $R_T$ es el radio de la Tierra y $\rho_T$ y $\rho_L$ son las densidades de la Tierra y la Luna, respectivamente.

El caso opuesto es considerar a la Luna como un fluido. En este caso, la Luna se deforma a medida que gira en torno a la Tierra, convirtiéndose en un cuerpo oblongo con cada revolución. En este caso, el límite de Roche es (Lowrie, 2007):

$$d_R = 2.42 R_T \left(\frac{\rho_T}{\rho_L}\right)^{1/3} \qquad (7)$$

Comparando (5) y (6), comprobamos que una Luna fluida se desintegraría al doble de distancia que una Luna totalmente rígida. En la realidad, la Luna no es completamente rígida, sino que presenta una cierta elasticidad. Por ello, se suele tomar como límite de Roche un valor intermedio entre los proporcionados por las expresiones (5) y (6).

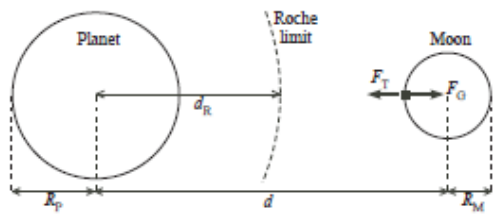 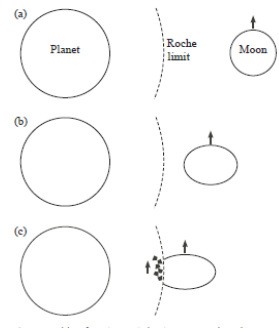

Fig. 4. Límite de Roche en el sistema Tierra-Luna para una Luna rígida (izquierda) y una Luna fluida (derecha). Imagen modificada de Lowrie (2007).

# 2. Datos observacionales

*"It's a literal hearth of darkness"*
Doyle

En el siguiente apartado, recogeré las evidencias observacionales del agujero negro supermasivo del centro de la Vía Láctea, así como los efectos más importantes de las mareas en el Sistema Solar.

## 2.1. El agujero negro supermasivo de la Vía Láctea

Como se ha comentado en el epígrafe 1.1, se cree que en el centro de cada galaxia existe un agujero negro supermasivo. El SMBH de nuestra galaxia, la Vía Láctea, se sitúa en dirección de la constelación de Sagitario, en una zona conocida como Sagitario A. Es una zona que no se puede observar en el óptico debido al alto contenido de polvo y la consecuente extinción, pero sí es observable en infrarrojo y radio. Los estudios de Andrea Ghez (Ghez et al., 1999) de seguimiento de las órbitas de estrellas en Sagitario A demostraron que existe una fuente de radio muy intensa en dicha región, denominada Sagitario A* (Figura 5), alrededor de la cual las estrellas cercanas giran en órbitas keplerianas. Ghez infirió una densidad para Sagitario A* de $10^{12} M_\odot/pc^3$, valor que sólo puede explicarse con la presencia de un SMBH en el centro de la Vía Láctea. Posteriormente, Schödel, siguiendo el método de Ghez, observó la estrella S2 orbitando Sagitario A* con un periodo de 15.2 años y un pericentro de 17 horas-luz. Este estudio le permitió llegar a la conclusión de que la masa alrededor de la cual orbitaba la estrella era de $4 \cdot 10^6 \, M_\odot$ (Schödel et al., 2002)

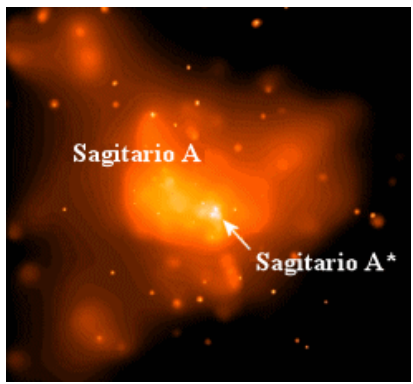

Fig. 5. Sagitario A y Sagitario A*, el centro de la Vía Láctea. Imagen del Observatorio de rayos-X Chandra (1999) vía http://ciencia.nasa.gov/science-at-nasa/2002/21feb_mwbh/

(Figura 6) concentrada en un radio menor de 0.001 pc, lo que respaldaba la hipótesis de Ghez.

Gracias a los estudios de Ghez y Schödel, la presencia de un agujero negro supermasivo en el centro de la Vía Láctea parece clara, aunque otra teoría sugiere la presencia de un cúmulo de estrellas jóvenes como alternativa al SMBH.

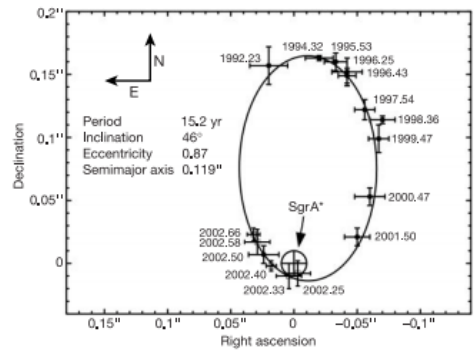

Fig. 6. órbita de la estrella S2 desde 1992 hasta 2002 alrededor de Sagitario A*, indicio de la presencia de un SMBH en el centro de la Vía Láctea. Imagen de Schödel (2002).

En la Vía Láctea se estima que existen unos 100 millones de agujeros negros de tipo estelar (agujeros negros menores), con masas del orden de 3 a 30 veces la masa del Sol. El agujero negro más próximo a la Tierra se sitúa a unos 300 años luz de nosotros, 100 veces más lejos que estrella más cercana al Sol, Próxima Centauri.

## 2.2. Mareas en el Sistema Solar

En este apartado analizaré el fenómeno de las mareas en aquellos objetos del Sistema Solar donde más se notan sus efectos, en concreto, estudiaré los casos de la Tierra, Ío, Europa y Encélado.

### 2.2.1. La Tierra

Como se ha explicado en el apartado 1.2, las mareas en la Tierra se producen por la atracción gravitatoria que existe entre la Luna y la Tierra, y, en menor medida, entre el Sol y la Tierra. Este efecto hace que la masa oceánica (y terrestre) se desplace de la posición de equilibrio siguiendo los armónicos esféricos de la Figura 3. De ellos, el efecto más importante es el semidiurno. Este armónico hace que se produzcan variaciones en la altura de la marea cada 12 horas.

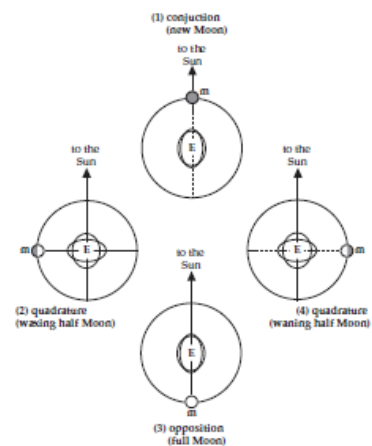

Fig. 7. Posición relativa de la Tierra, la Luna y el Sol, causantes de las *spring* y *neap tides*. Imagen de Lowrie (2007).

Existen dos casos especialmente interesantes que tiene que ver con la posición relativa del Sol, la Luna y la Tierra sobre la eclíptica (Figura 7). Para simplificar la explicación, consideraré que los tres astros se encuentran en el mismo plano (en realidad no es así, ya que la órbita de la Luna está inclinada 5º respecto de la eclíptica):

- *Spring tides*: se producen cuando los tres astros se encuentran o bien en conjunción o en oposición. Es el momento en el que la altura de la marea alcanza su máximo.

- *Neap tides*: es el caso contrario al anterior. Se producen cuando la Tierra, el Sol y la Luna forman un ángulo recto (cuadratura) que hace que la altura de la marea sea la mínima.

Las mareas más altas de la Tierra (excluyendo las conjunciones y las oposiciones) las encontramos en la bahía de Mont Saint-Michel (Francia) (Figura 8) y en la bahía de Fundy (Canadá), con unas alturas máximas de 14.5 metros y 17.6 metros, respectivamente (datos vía http://www.tablademareas.com/).

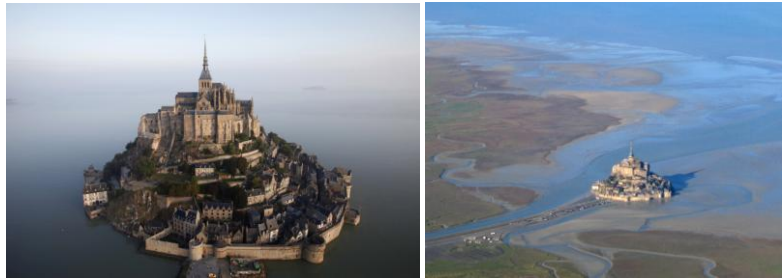

Fig. 8. Mont Saint-Michel en presencia de marea alta (*izda.*) y marea baja (*drcha.)*. Imágenes vía Google Images.

### 2.2.2. Ío

Ío es la luna más interna de Júpiter, descubierta por Galileo en 1610. El efecto de las mareas en Ío no se debe sólo a Júpiter, sino que también intervienen los otros satélites galileanos: Europa, Ganímedes y Calisto (de dentro hacia fuera). Como las velocidades de traslación de las lunas son diferentes, cada una tira de Ío en una dirección determinada hacia fuera del sistema. En cambio, la presencia de Júpiter hace que tire de Ío hacia el interior (Figura 9). Este efecto de "tira y afloja" por parte de los cuatro astros hacen que Ío presente un movimiento de bamboleo según completa su órbita. Debido a estas interacciones, en Ío se producen grandes mareas terrestres (carece de agua en superficie). La gran energía gravitacional y rotacional que Ío almacena debido a este encogimiento y estiramiento del satélite se transforma en calor. Esta acumulación de calor hace que la temperatura del interior de Ío aumente hasta el punto de fundir parte del material interno. Para que el sistema se equilibre, ese calor acumulado debe salir al exterior, y lo hace de dos formas: como eyecciones volcánicas de sulfuros y $SO_2$ y como flujos de lava enriquecidos en magnesio y hierro, que alcanzan una altura de 100 metros sobre la superficie. A este proceso de transformación de energía gravitacional y rotacional en calor se le denomina calentamiento por fuerzas de marea (Peale et al. 1979)*.*

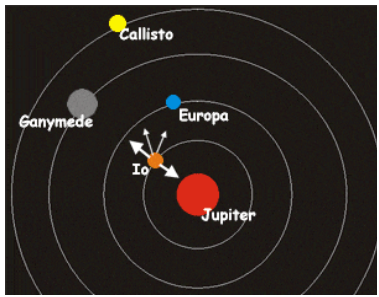

Fig. 9. La acción conjunta de Júpiter y el resto de satélites hacen que Ío presente un movimiento de bamboleo, responsable de su actividad geológica. Imagen de la NASA (2015) vía http://spaceplace.nasa.gov/io-tides/sp/

Se cuentan alrededor de unos 300 volcanes activos en Ío, con unas velocidades de salida de flujos de unos 500 m/s. Algunas de ellas han sido captadas por las sondas Galileo y Voyager 1 de la NASA. Debido a la gran actividad volcánica en Ío, el satélite no presenta ningún cráter de impacto, ya que el vulcanismo ha eliminado toda huella de cualquier meteorito.

### 2.2.3. Otros satélites

El efecto de calentamiento por fuerzas de marea también se da en otras lunas del Sistema Solar, como Europa o Encélado. En Europa (segundo satélite de Júpiter desde dentro), el efecto de calentamiento por fuerzas de marea se traduce en una actividad volcánica dominada por el criovulcanismo, en el que los volcanes expulsan flujos de hielo. Esto hace que, al llegar a la superficie, estos flujos de hielo se solidifiquen y hagan de la superficie de Europa una superficie muy lisa cubierta de hielo.

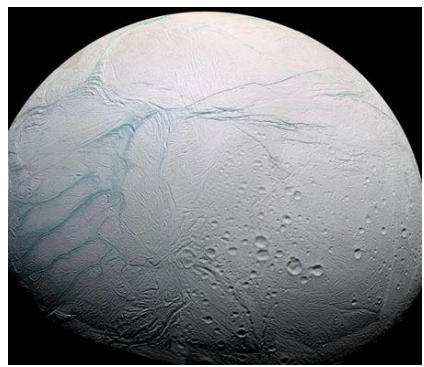

Fig. 10. La superficie de Encélado está totalmente cubierta de hielo. Imagen Cassini Imaging Team (2005) vía http://apod.nasa.gov/apod/ap050906.html

Algo similar ocurre en Encélado, uno de los satélites de Saturno. Al igual que en Europa, el criovulcanismo es la causa de que su superficie esté cubierta de hielo (Figura 10), aunque el origen de su actividad geológica no está del todo claro.

# 3. *Interstellar*: análisis y resultados

*"It's not a ghost. It's gravity"*
Cooper

Una vez que hemos estudiado los conceptos de agujeros negros y mareas en la teoría y con aplicación a la Vía Láctea y al Sistema Solar, respectivamente, paso a introducirme en el universo de *Interstellar*.

## 3.1. Gargantúa

Cuando el profesor Brand le explica a Cooper su plan para salvar a la humanidad, éste le responde "*No hay planeta en el Sistema Solar capaz de albergar vida*". El profesor Brand asiente y le explica que cerca de Saturno se ha descubierto un agujero de gusano que comunica nuestro Sistema Solar con el centro de otra galaxia y que, en esa nueva galaxia, existen tres planetas potencialmente habitables. Por tanto, la misión está clara: es

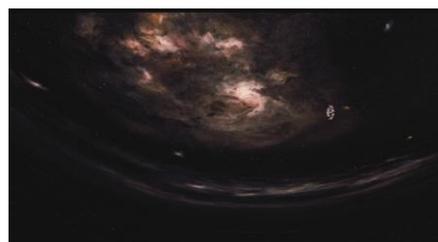

Fig. 11. La nave *Endurance* a punto de cruzar el agujero de gusano. Imagen de *Interstellar* (2014).

necesario cruzar el agujero de gusano y llegar a la nueva galaxia (Figura 11).

Una vez que la *Endurance* llega a su destino, todo está en calma. La tripulación se encuentra en el oscuro mar del vacío, tan sólo roto por las estrellas de fondo, algunas nebulosas de emisión y un objeto que domina sobre los demás: Gargantúa*.*

La primera imagen de Gargantúa puede sorprender, ya que su aspecto es el que esperaríamos de una estrella corriente, como nuestro Sol. Eso es debido al disco de acreción que le rodea (Figura 12). Los discos de acreción alrededor de SMBH se generan cuando una estrella pasa muy cerca de ellos. El intenso campo gravitatorio del SMBH destroza a la estrella y parte del gas remanente es capturado, formándose así el disco de acreción. Una simulación de este proceso, llevada a cabo por Suvi Gezari (Universidad John Hopkins) y James Guillochon (UCLA) puede verse en el siguiente enlace: http://hubblesite.org/newscenter/archive/releases/2012/18/video/a/. En ella se aprecia como una estrella próxima a un SMBH se desintegra por completo en menos de 140 días desde el contacto, quedando únicamente el disco de acreción y una cola de material que escapa del campo gravitatorio generado por el agujero negro.

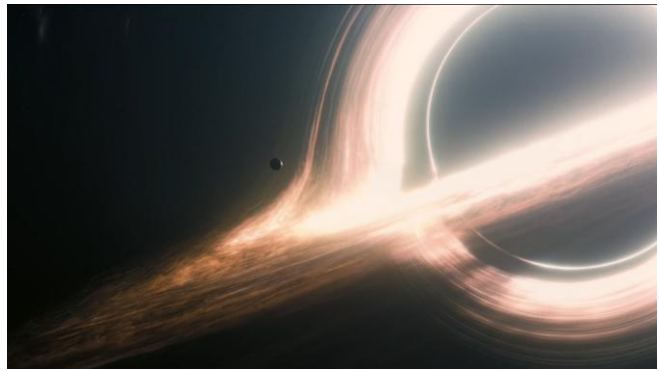

Fig. 12. Gargantúa rodeado del disco de acreción. Se observa al planeta de Miller orbitándolo. Imagen de *Interstellar* (2014).

Los discos de acreción suelen tener unas temperaturas de millones de Kelvin y son una gran fuente de radiación gamma y X. Para evitar que la tripulación de la *Endurance* tuviese problemas cerca de Gargantúa, Christopher Nolan y Kip Thorne idearon un disco de acreción con una temperatura próxima a la superficie del Sol, unos 6000 K (Thorne, 2014).

Así, la zona próxima a Gargantúa puede asimilarse a la vecindad del Sol, por lo menos en cuanto a radiación se refiere, con el máximo de visión en el rango del visible siguiendo la ley de Wien. Para que un disco de acreción llegue a esas temperaturas tan "bajas" es necesario que el SMBH no haya absorbido materia en mucho tiempo, evitando así el calentamiento del disco. De esta manera, la tripulación de la *Endurance* estaría a salvo de la radiación de Gargantúa.

La imagen más impresionante de Gargantúa la encontramos a medida que la *Endurance* se acerca a él. En ella, vemos una prominente esfera negra rodeada de un disco de acreción en su plano ecuatorial pero también en su plano polar, una imagen muy diferente a las imágenes a las que estamos acostumbrados de agujeros negros. La diferencia radica en que *Interstellar* es la primera película que tiene en cuenta el efecto de lente gravitacional. La luz del disco de acreción de detrás de Gargantúa llega al espectador por encima del agujero negro debido a su extremo campo gravitatorio, que hace que la trayectoria de la luz se curve. La Figura 13 muestra dos imágenes de un SMBH en dos películas distintas: en la primera, perteneciente a la película *The Black Hole* no se tiene en cuenta el efecto de lente gravitacional mientras que en la segunda, de *Interstellar*, sí tiene en cuenta este efecto. Este hecho ha sido aclamado por la comunidad científica por ser la primera vez que se presenta un agujero negro supermasivo con disco de acreción tal y como se vería en la realidad.

Una explicación más detallada del efecto de lente gravitacional en agujeros negros puede verse en Thorne (2014), así como el proceso de construcción de Gargantúa que llevó a cabo el equipo *Double Negative*, encargado de la parte gráfica de la película.

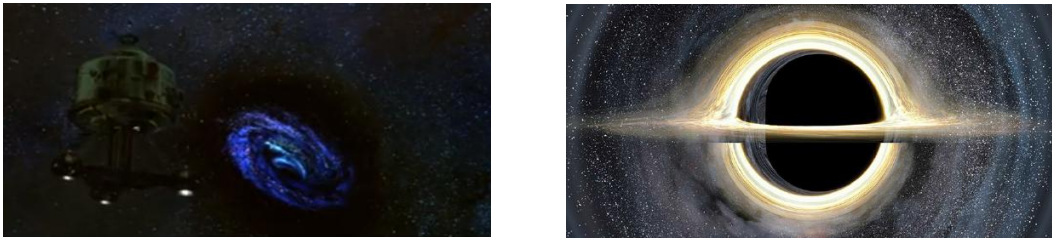

Fig. 13. A la izquierda, imagen de un agujero negro supermasivo de la película *The Black Hole* (1979). A la derecha, imagen de *Interstellar* (2014). La diferencia radica en el efecto de *lente gravitacional.*

### 3.2. Planeta de Miller

Tras observar Gargantúa, la tripulación de la *Endurance* debe continuar con su misión. El primer planeta que estudian Cooper y sus compañeros es el planeta de Miller, un planeta cubierto totalmente por agua que se encuentra peligrosamente cerca del horizonte de eventos de Gargantúa (Figura 14).

Para que el planeta de Miller pueda orbitar alrededor de Gargantúa, debe cumplir dos condiciones:

- Su radio orbital debe ser superior al radio de Schwarzschild de Gargantúa.

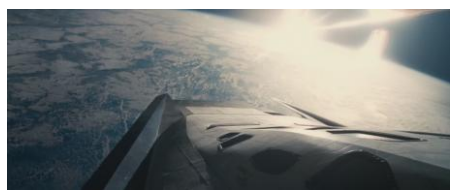

Fig. 14. El planeta de Miller, totalmente cubierto de agua, desde la nave auxiliar *Ranger*. Al fondo puede verse *Gargantúa*. Imagen de *Interstellar* (2014).

- El límite de Roche de Gargantúa debe ser inferior al radio orbital del planeta de Miller.

La Tabla 1 muestra los resultados obtenidos del radio de Schwarzschild y el límite de Roche para tres SMBH reales y un hipotético planeta que los orbita con una densidad igual a la de la Tierra. Se ha considerado como densidad del agujero negro supermasivo la obtenida de dividir su masa asociada entre el volumen definido por su radio de Schwarzschild.

### TABLA 1. PARÁMETROS ORBITALES

| Galaxia | $M_\bullet$ ($M_\odot$) | $R_{Sch}$ (m) | $\rho_\bullet$ (kg·m$^{-3}$) | $d_R$ (m) |
|---|---|---|---|---|
| M32 | $3.9·10^6$ | $1.15·10^{10}$ | $1.22·10^6$ | $1.39·10^{11}$ |
| NGC 2778 | $1.3·10^7$ | $3.83·10^{10}$ | $1.10·10^5$ | $2.08·10^{11}$ |
| NGC 5845 | $2.9·10^8$ | $8.55·10^{11}$ | 220.35 | $5.85·10^{11}$ |

$M_\bullet$ = masa del SMBH (en masas solares). Datos de las masas: Kormendy y Gebhardt (2001)

$R_{Sch}$: Radio de Schwarzschild del SMBH obtenido a partir de (1)

$\rho_\bullet$: Densidad del SMBH.

$d_R$: límite de Roche obtenido con una expresión intermedia entre (5) y (6): $d_R = 2 \cdot R_{Sch} \left(\frac{\rho_\bullet}{\rho_{Miller}}\right)^{1/3}$

$M_\odot$ = $1.989·10^{30}$ kg, $\rho_{Tierra}$ = 5514 kg·m$^{-3}$, c = $3·10^8$ m·s$^{-1}$, G = $6.67·10^{-11}$ N·m$^{-2}$·kg$^{-2}$.

Si Gargantúa fuese uno de los agujeros negros supermasivos recogidos en la Tabla 1, el planeta de Miller, si tuviera la misma densidad que la Tierra, debería estar a una distancia superior, tanto del límite de Roche como del radio de Schwarzschild. De aquí en adelante, consideraré que Gargantúa posee la misma masa que NGC 2778, es decir, $1.3·10^7$ $M_\odot$ y que el planeta de Miller se sitúa a una distancia de Gargantúa de $3·10^{11}$ metros en una órbita circular

Aún así, el planeta de Miller no estaría salvado del todo, ya que, debido al tirón gravitacional del agujero negro, el planeta podría salir despedido de su órbita. Para que esto no ocurra, el planeta debe situarse en una órbita estable alrededor del SMBH, y esto sucede, en la versión más simple, cuando la fuerza gravitatoria que empuja al planeta hacia el agujero negro se equilibra con la fuerza centrífuga que tiende a alejar al planeta. Suponiendo que el planeta de Miller sigue una órbita circular, se debe cumplir que la fuerza gravitatoria compense a la fuerza centrífuga, es decir:

$$\frac{GMm}{r^2} = \frac{mv^2}{r} \qquad (8)$$

en donde M es la masa del SMBH, m es la masa del planeta, v es su velocidad de traslación y r es la distancia que separa ambos cuerpos. La velocidad a la que la órbita es estable es, por tanto:

$$v = \sqrt{\frac{GM}{r}} \qquad (9)$$

Aplicando la expresión (8) para la masa de Gargantúa y la distancia orbital del planeta de Miller considerados, obtenemos que la velocidad orbital con la que el planeta de Miller se desplaza alrededor de Gargantúa es de 7.58·10$^7$ m·s$^{-1}$ (0.25 veces la velocidad de la luz). Esto significa que el planeta de Miller orbitaría Gargantúa una vez cada 7 horas.

Por tanto, un planeta en una órbita circular y con la densidad de la Tierra sí podría orbitar alrededor de un agujero negro supermasivo. En *Interstellar,* la única referencia a la órbita del planeta de Miller la encontramos en uno de los diálogos, en el que Romilly les explica a sus compañeros que el planeta de Miller se sitúa alrededor de Gargantúa "*como una pelota alrededor de un aro de baloncesto*". Con esto, Romilly quiere expresar lo cerca que están del agujero negro y lo peligroso que resulta situarse allí, debido a la dilatación gravitacional del tiempo que sufrirían. Por ello, la *Endurance* deberá situarse en una órbita mucho mayor que la del planeta y, así, minimizar lo máximo posible la dilatación temporal (Figura 15).

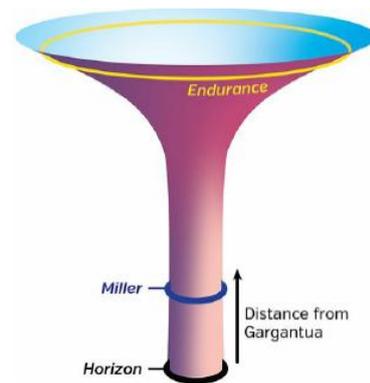

Fig. 15. Esquema de la órbita del planeta de Miller (azul) alrededor de Gargantúa y de la nave *Endurance* (amarillo). Imagen de Thorne (2014).

### 3.2.1. Dilatación del tiempo en el planeta de Miller

La cercanía del planeta de Miller a Gargantúa es la responsable de la dilatación temporal que se menciona en la película. En ella, se cita que una hora en la superficie del planeta de Miller equivaldrían a siete años en la Tierra.

Si consideramos que Gargantúa es un SMBH esférico y no rotatorio, la dilatación del tiempo en el planeta de Miller vendrá dada por la expresión (2). El cociente entre los dos intervalos de tiempo ($\Delta t/\Delta t_0$) deberá ser del orden 6·10$^4$ para que una hora en el planeta correspondan a siete años (61320 horas) en la Tierra.

En nuestro caso, con una masa de Gargantúa igual a la de NGC 2778 y un radio orbital del planeta de Miller de 3·10$^{11}$ metros, el cociente ($\Delta t/\Delta t_0$) es de 1.07, o lo que es lo mismo, una hora en el planeta de Miller corresponderían a 1.07 horas en la Tierra. Claramente, la dilatación temporal es muy inferior a la deseada. Para que nuestro planeta tuviese semejante dilatación temporal, debería situarse a una distancia de 10 centímetros superior al radio de Schwarzschild de Gargantúa. A esta distancia, el planeta sí sufriría una dilatación temporal semejante a la de *Interstellar*, pero estaría muy por debajo del límite de Roche, por lo que el planeta se habría desintegrado antes de llegar a la distancia requerida. Por tanto, un agujero negro supermasivo esférico y no rotatorio no puede explicar la dilatación temporal que se menciona en *Interstellar*.

Aún así, cabe mencionar que existe dicha dilatación. La máxima dilatación gravitacional temporal que observaríamos en el planeta de Miller antes de llegar al límite de Roche de Gargantúa, sería de un factor 1.1, es decir, una hora en la superficie del planeta de Miller correspondería a 1.1 horas en la Tierra.

### 3.2.2. Mareas en el planeta de Miller: planeta rígido y deformable

Una vez que hemos visto la estabilidad de la órbita de un planeta alrededor de un agujero negro y la dilatación temporal que sufre, podemos analizar el fenómeno de las olas en el planeta de Miller. Para este apartado, asumiré que las olas del planeta se comportan como mareas (de hecho, se le llama ola a una columna de agua de poca altura y baja masa que, debido a la interacción con la atmósfera, se desplaza, mientras que el fenómeno de las mareas implica alturas de columnas y masas de agua mayores que sienten la interacción gravitatoria) y que en el planeta de Miller, las mareas terrestres son despreciables. Distinguiré entre un planeta de Miller rígido y deformable.

*Planeta rígido*

Podemos calcular la altura que alcanzaría la marea en el planeta de Miller a partir de la expresión (4). Para ello, consideraré que la masa de Gargantúa es la del apartado anterior y un radio del planeta de Miller igual al radio de la Tierra. En *Interstellar* se especifica una gravedad en el planeta de Miller de 1.3 veces la gravedad de la Tierra, es decir, $g_{Miller} = 12.74$ m·s$^{-2}$. Los resultados, para cinco latitudes diferentes, se recogen en la Tabla 2.

**TABLA 2. ALTURA DE LA MAREA**

| Latitud (º) | Potencial de marea (J·kg$^{-1}$) | Altura de la marea (m) |
|---|---|---|
| 90 | -1.30·10$^6$ | -1.02·10$^5$ |
| 60 | -3.24·10$^5$ | -2.54·10$^4$ |
| 45 | 6.48·10$^5$ | 5.09·10$^4$ |
| 30 | 1.62·10$^6$ | 1.27·10$^5$ |
| 0 | 2.59·10$^6$ | 2.03·10$^5$ |

Latitudes correspondientes al Hemisferio Norte. Nótese que a latitudes altas, la altura de la marea es negativa, por lo que el nivel teórico del agua estaría por debajo de la superficie.

Estos datos explican la gran altura que alcanzan las olas en el planeta de Miller, como la que se muestra en la Figura 16.

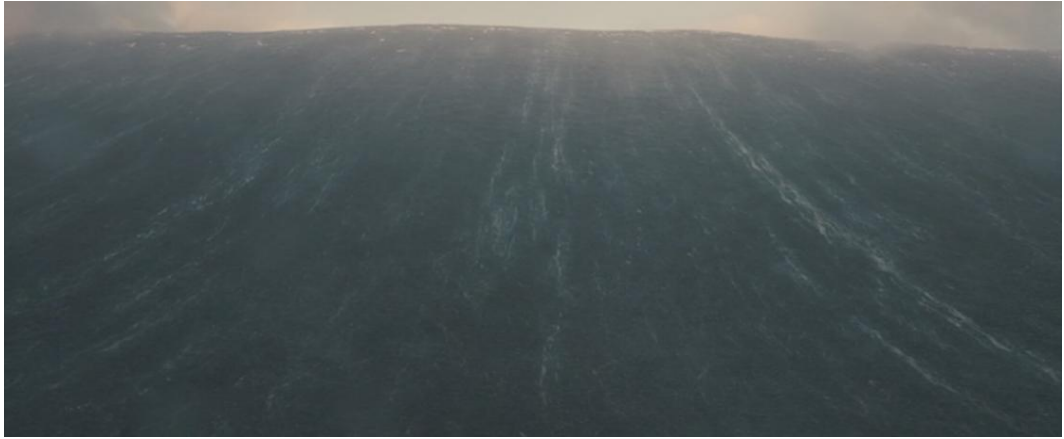

Fig. 16. Una de las enormes olas del planeta de Miller, causada por su proximidad a Gargantúa. Imagen de *Interstellar* (2014).

A continuación, una vez que hemos visto que es posible encontrar mareas tan altas como las observadas en *Interstellar,* me centraré en la periodicidad con las que éstas se observan.

La forma más sencilla de estudiar la periodicidad de las olas en un planeta rígido es comparar la velocidad de rotación del planeta con su velocidad de traslación. Según esto, podemos distinguir los casos en los que el planeta rota más rápido de lo que se desplaza (su velocidad de rotación sería mayor que 0.25 veces la velocidad de la luz) o que la velocidad de traslación fuese igual a la de rotación.

- <u>Planeta con rotación rápida</u>: es el caso explicado en el epígrafe 1.2. Asumiendo un potencial de marea con la forma de (2), el término dominante correspondería a una periodicidad semidiurna, es decir, una variación de la marea cada 12 horas. En *Interstellar*, el tiempo que transcurre entre marea y marea (o entre ola y ola) es mucho menor. Para que la periodicidad semidiurna fuese la responsable de las mareas del planeta de Miller, se debería hacer de noche entre marea alta y marea alta, algo que no se refleja en la película.
  Sin embargo, sí que es posible explicar las mareas de la película con un potencial de marea, aunque introduciendo matices. Como ya se ha comentado, la expresión (2) es una primera aproximación al potencial de marea. Si quisiéramos ver mareas con mayor frecuencia en el planeta, habría que extender el desarrollo del potencial de marea general hasta órdenes mayores. Así, llegaríamos a una expresión que, al descomponerla en sumandos según su periodicidad, reflejaría las mareas observadas. Si consideramos un periodo entre marea y marea de 1 hora, sería necesario desarrollar el potencial hasta obtener un término que dependiese de $\cos(6H)$. El desarrollo del potencial de marea en órdenes mayores puede verse en Melchior (1966).

- <u>Planeta con rotación lenta</u>: el caso más sencillo es considerar que la velocidad de traslación es igual a la velocidad de rotación. En esta hipótesis, el planeta de Miller estaría siempre dándole la misma cara a Gargantúa, en cuyo caso no habría una sucesión de mareas, sino que existiría un aumento en la altura de la masa oceánica en dirección a Gargantúa que se mantendría fijo (Figura 17). Un posible mecanismo que explique las olas en este caso podría venir dado por la 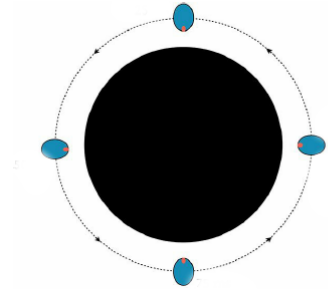

Fig. 17. El planeta de Miller dando la misma cara a Gargantúa. Imagen modificada de Thorne (2014).

fricción de la marea. Si el potencial generado por Gargantúa posee una variación temporal armónica, se produciría un par de fuerzas que tenderían a colocar al planeta de Miller en una posición estable. Este movimiento de bamboleo producido por el par de fuerzas (Figura 18) haría que la masa oceánica tuviera un movimiento de vaivén, que sería el responsable de las mareas.

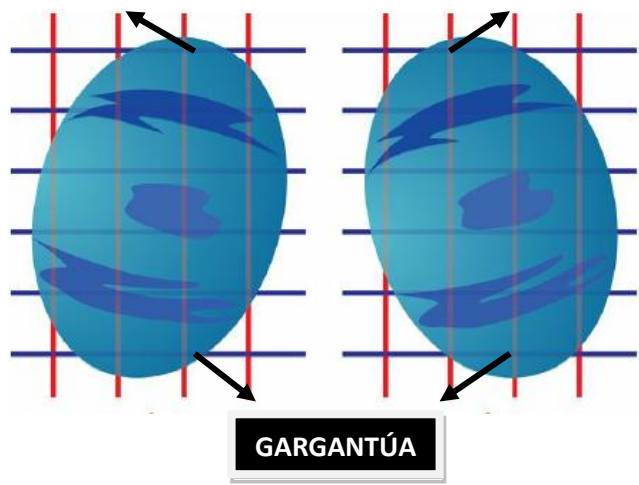

Fig. 18. Un agujero negro pulsante haría que el planeta de Miller presentase un par de fuerzas (flechas negras) que provocarían en el planeta un movimiento de vaivén. Imagen modificada de Thorne (2014).

Otra posible explicación al movimiento de bamboleo causante de la marea podría deberse a la interacción del planeta de Miller con, además de Gargantúa, los otros planetas vecinos, el planeta de Mann y el planeta de Edmunds. De forma análoga a Ío, las interacciones entre los tres cuerpos puede hacer que el planeta de Miller presente este movimiento.

Aún así, el bamboleo no es capaz de explicar lo observado en *Interstellar*, ya que, en la película, se observa cómo dos olas diferentes golpean a la tripulación en poco tiempo, mientras que, con esta explicación, sería la misma ola la que impactase contra ellos.

*Planeta deformable*

Considerar el carácter deformable del planeta de Miller simplifica notablemente, al menos en términos numéricos, la explicación de las olas en *Interstellar*. A continuación se presentan dos posibles mecanismos por los que se pueden generar olas de gran altura en el planeta de Miller.

El primero de ellos consiste en estudiar la respuesta de la litosfera al paso de una marea. Una masa de agua tan pesada puede hundir la litosfera a su paso. Cuanto más pesada sea la masa de agua, más tenderá a flexionarse la superficie por debajo. Idealizando la marea como un triángulo (Figura 19), es claro ver que la máxima flexión de la litosfera se sitúa en la dirección vertical del máximo de altura de la marea. A medida que la marea avanza, el máximo de altura se va desplazando con ella, seguido de la flexión de la litosfera. Por tanto, la flexión se desplaza sobre la litosfera con la marea.

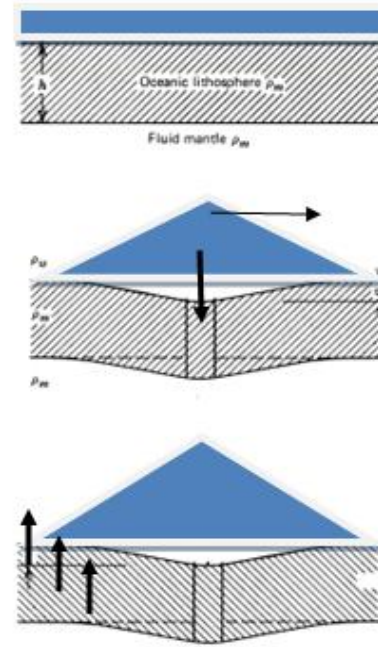

Una vez que el máximo de masa de agua ha pasado por la litosfera que ha deformado, ésta tiende a elevarse para contrarrestar su hundimiento. Esta respuesta de la litosfera empujaría el agua que tiende por encima hacia arriba, sumándose así a la marea. El efecto se retroalimenta cada vez que la litosfera se hunde y vuelve a levantarse, por lo que la marea resultante puede alcanzar grandes alturas.

Fig. 19. Aumento de la altura y la velocidad de la marea por flexión y respuesta de la litosfera. Imagen modificada de Turcotte (2002).

La elevación de la litosfera también puede ser debida a otro factor. Si consideramos que el planeta de Miller posee un núcleo externo fluido como el de la Tierra, las fuerzas de marea también tenderán deformar el núcleo. Así, cuando el núcleo del planeta de Miller se deforme debido a las fuerzas de marea de Gargantúa, tenderá a levantar el material que tiene por encima, y, con ello, a la litosfera. Este efecto se sumará al proceso de la Figura 19.

La segunda explicación se debe a la existencia de terremotos en el planeta. La energía acumulada en el interior del planeta de Miller, debida las interacciones con otros planetas vecinos (planeta de Mann y planeta de Edmunds) como en el caso de Ío, o debida a una teórica tectónica de placas que tuviese lugar en su interior, se libera en forma de ondas sísmicas. A esa liberación de energía se le denomina terremoto (en el planeta de Miller sería más correcto denominarlos "millermotos"). Si uno de estos millermotos ocurre en una zona oceánica (sería lo más probable, ya que en la película en ningún caso se menciona que exista superficie sin agua en el planeta de Miller) se denominan tsunamis. La Figura 20 muestra cómo se genera uno de estos tsunamis.

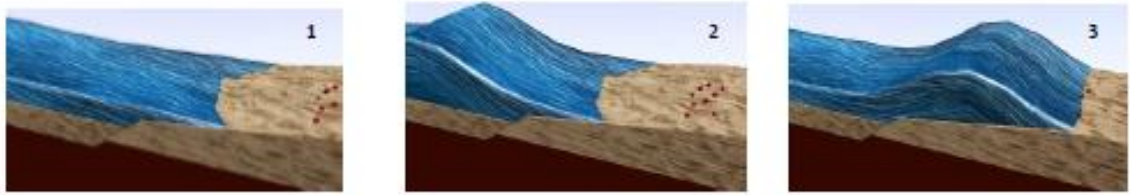

Fig. 20. Formación de un tsunami. 1- Ruptura de la litosfera. 2-Generación de la ola debida la fractura. 3-Propagación de la ola. Simulación de Carmen Álvarez Cobos (Facultad de Ciencias del Mar y Ambientales, Universidad de Cádiz). Imágenes vía https://www.youtube.com/watch?v=HMksNCzVXE8

Cuando la ola generada en aguas profundas se acerca a la costa y la profundidad de las aguas disminuye, la ola se ralentiza. La ola pierde energía cinética, la cual se transforma en energía potencial o se pierde debido al rozamiento con el suelo. Esto provoca un aumento en la altura de la ola.

Si se produce más de un tsunami, al llegar a la costa, el intervalo de tiempo entre ola y ola disminuye. La ilustración de este proceso puede verse en la Figura 21.

La velocidad a la que se propaga el tsunami en aguas profundas se puede calcular a partir de la siguiente expresión (Lowrie, 2007):

$$v = \sqrt{D \cdot g}$$

( 10 )

en donde g es la gravedad del planeta y D es la profundidad a la que se produce el seísmo. La Tabla 3 muestra distintas velocidades de tsunamis considerando varias profundidades del seísmo en el planeta de Miller.

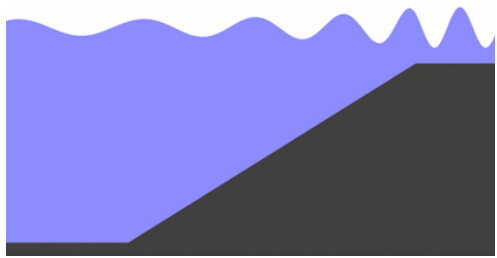

Fig. 21. El tsunami originado en aguas profundas se convierte en olas de gran altura y frecuencia. Imagen de Regis Lachaume (2015) vía https://es.wikipedia.org/wiki/Tsunami#/media/File:Propagation_du_tsunami_en_profondeur_variable.gif

**TABLA 3. VELOCIDADES DE TSUNAMIS**

| D (m) | v (m/s) |
|---|---|
| **1000** | 112,87 |
| **3000** | 195,50 |
| **5000** | 252,39 |
| **8000** | 319,25 |
| **10000** | 356,93 |

D=profundidad del seísmo

v=velocidad del tsunami obtenida a partir de (9) con $g_{Miller} = 12.74 \, m \cdot s^{-2}$.

Hemos obtenido una forma de producir olas de gran altura y que presenten una alta frecuencia, justo lo que necesitábamos para reproducir las olas observadas en el planeta de Miller. En mi interpretación de la película, se produce una serie de "millermotos" de magnitud muy alta en aguas muy profundas en el planeta de Miller (nótese que a mayor profundidad del seísmo, mayor es su velocidad de propagación). Esos "millermotos" generan tsunamis de gran altura y baja frecuencia, que se adentran en las aguas poco profundas. Aquí, las olas se hacen más altas y el tiempo entre ola y ola es menor. La nave auxiliar *Ranger* "amilleriza" precisamente en una situación de "entre-olas" en aguas poco profundas (esta hipótesis se comprueba en la película, ya que se ve cómo la tripulación abandona la *Ranger* y comienza a andar sobre la superficie del planeta, en donde el agua les cubre escasamente por debajo de las rodillas). Entonces, Amelia observa cómo una de las olas que acaba de pasar justo antes de su "amillerizaje" se aleja de ellos. En este momento, Cooper le advierte que otra ola los va a golpear por detrás, por lo que deben regresar a la nave. Esta escena es en una de las escenas más espectaculares pero a la vez más angustiosas de *Interstellar*, porque se ve a Amelia luchando por llegar a la nave antes de que la ola impacte contra ellos.

Una vez que la *Ranger* logra abandonar el planeta de Miller, ponen rumbo hacia el siguiente planeta, el planeta de Mann.

### 3.3. Planeta de Mann

El planeta de Mann es un planeta inhóspito y totalmente cubierto de hielo en la superficie, como puede verse en la Figura 22.

Una primera explicación a este hecho correspondería a la presencia de criovulcanismo en el planeta de Mann. Al igual que sucede en el caso de Europa, la acción gravitatoria combinada de Gargantúa y el planeta de Miller pueden hacer que sobre el planeta de Mann se activen fuerzas internas que tiendan a expulsar hielo a la superficie.

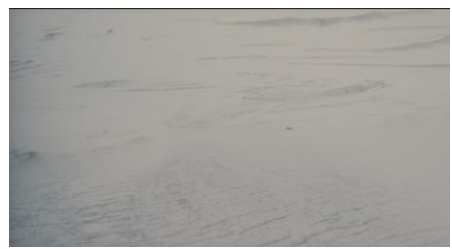

Fig. 22. Superficie del planeta de Mann, completamente helada. Imagen de *Interstellar* (2014).

Sin embargo, existe otra teoría, que tiene más que ver con la Tierra, por la que un planeta como el de Mann podría estar totalmente cubierto de hielo. Esta teoría se denomina *Snowball Earth* o *Tierra bola de nieve.* Según la teoría de *Snowball Earth*, la Tierra quedó completamente cubierta por hielo hace unos 650 millones de años, en un periodo llamado Criógeno, en la que la temperatura de la Tierra cayó hasta los -50ºC. La principal prueba a favor de esta teoría fue el descubrimiento de till glacial en latitudes bajas. El till glacial es un conglomerado de sedimentos producto del transporte glacial y característico de eras glaciales. La única explicación posible a encontrar estos tills es latitudes bajas es considerar que los casquetes polares avanzaron hasta estas latitudes y, a medida que el hielo se fue derritiendo y los casquetes fueron retrocediendo, dejaron tras de sí estos tills. La Figura 23 muestra los puntos de la Tierra en los que se han hallado restos de tills. El hallazgo que supuso un gran punto de apoyo para esta teoría se realizó en 1964, a cargo del geólogo W. Brian Harland, quien descubrió tills en Svalbard (Islandia) y Groenlandia con ciertas características de zonas tropicales (Harland et al. 1964).

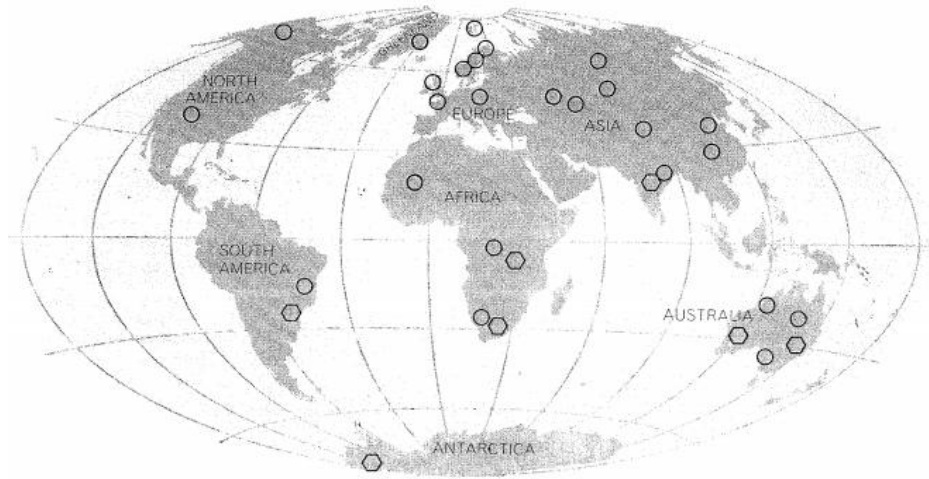

Fig. 23. Distribución actual de tills de la era precámbrica (círculos) y del carbonífero (hexágonos). Esta distribución supone una gran prueba de la presencia de casquetes polares en latitudes bajas. Imagen de Harland et al. (1964).

Otra evidencia importante de la *Snowball Earth* viene dada por el análisis isotópico del $^{13}$C. Este isótopo del carbono es un isótopo estable que está relacionado con la actividad biológica de los seres vivos. Representa en la actualidad el 1.1% del carbono terrestre. Los estudios de Hoffman en rocas del cratón del Congo (Namibia) indicaban una anomalía negativa en la concentración del $^{13}$C. Hoffman concluyó que dicha anomalía sólo podía explicarse debido a la desaparición de la actividad biológica provocada por una era glacial. Este periodo terminó drásticamente debido a la actividad volcánica, que expulsó a la atmósfera gran cantidad de $CO_2$, lo que produjo un calentamiento del planeta (Hoffman et al. 1998).

Aunque la producción biológica se viese muy mermada debido a la era glacial y al drástico enfriamiento del planeta, el descubrimiento de microorganismo extremófilos asociados a ambientes carentes de luz solar, como pueden ser las fumarolas negras en las dorsales oceánicas, supone que la vida continuó existiendo y que superó la era glacial.

En *Interstellar*, el planeta de Mann puede estar atravesando una de estos periodos de *Snowball*.

# 4. Conclusiones

> *"Accident is the first building block of evolution"*
> Amelia Brand

El estudio realizado nos permite obtener las siguientes conclusiones:

1. El agujero negro supermasivo, Gargantúa, es la mejor recreación que jamás se ha hecho de un agujero negro, considerando, entre otros factores, el efecto de lente gravitacional que actúa sobre su disco de acreción.

2. Queda demostrado que un planeta con la densidad de la Tierra puede orbitar un agujero negro supermasivo si su distancia a éste es mayor que el límite de Roche y el radio de Schwarzschild del agujero negro.

3. La dilatación gravitacional del tiempo observada en el planeta de Miller no es compatible con el modelo de un agujero negro no rotatorio, ya que existe una gran discrepancia entre el valor establecido en la película y el valor obtenido en este trabajo. El resultado podrá mejorarse considerando otros tipos de agujeros negros, como los agujeros negros rotatorios.

4. A su vez, se ha demostrado que es posible encontrar olas del tamaño de las observadas en la película, a partir del estudio del potencial de marea. Los cálculos sugieren una altura máxima de la marea en el ecuador de unos 200 km, aunque la periodicidad de las olas no se puede explicar por este fenómeno.

5. La mejor explicación al fenómeno de las olas viene de considerar tsunamis en aguas muy profundas, los cuales avanzan hacia aguas poco profundas aumentando su altura y frecuencia.

6. En cuanto al planeta de Mann, hemos visto cómo un planeta puede presentar una superficie totalmente helada a partir de dos teorías: el criovulcanismo y la *Snowball Earth*.

Debido a esto, podemos establecer que *Interstellar* cumple razonablemente bien las leyes de la Física enunciadas en este trabajo. Las discrepancias con la realidad pueden ser debidas a licencias de guión por parte del director, con el fin de dotar a la película de un mayor dramatismo y espectacularidad.

# 5. Referencias

## Otra bibliografía consultada y recursos de interés